\documentclass[12pt,preprint]{aastex}

\usepackage{bm}
\usepackage{amsmath}
\usepackage{graphicx}
\usepackage{epstopdf}
\usepackage{color}

\begin{document}

\title{Magnetohydrodynamic Simulations for Studying Solar Flare Trigger Mechanism  \\
   }

\author{J. Muhamad \altaffilmark{1}, K. Kusano, S. Inoue \altaffilmark{2}, D. Shiota\altaffilmark{3}}
\affil{Institute for Space-Earth Environmental Research, Nagoya University, Furocho, Chikusa-ku, Nagoya, Aichi, 464-8601, Japan}

\altaffiltext{1}{Space Science Center, National Institute of Aeronautics and Space (LAPAN), 
Jl. Djundjunan 133, Bandung, 40173, Indonesia}
\altaffiltext{2}{Max Planck Institute for Solar System Research, G\"{o}ttingen, Germany }
\altaffiltext{3}{National Institute of Information and Communications Technology, 4-2-1
Nukui-Kitamachi, Koganei, Tokyo 184-8795, Japan}

\begin{abstract}
In order to understand the flare trigger mechanism, we conducted three-dimensional magnetohydrodynamic simulations using a coronal magnetic field model derived from data observed by the Hinode satellite. 
Several types of magnetic bipoles were imposed into the photospheric boundary of the Non-linear Force-Free Field (NLFFF) model of Active Region NOAA 10930 on 2006 December 13 to investigate what kind of magnetic disturbance may trigger the flare. 
As a result, we confirm that certain small bipole fields, which emerge into the highly sheared global magnetic field of an active region, 
can effectively trigger a flare. These bipole fields can be classified into two groups based on their orientation 
relative to the polarity inversion line: the so called opposite polarity (OP) and reversed shear (RS) structures as it was suggested by \citet{kus12}. 
We also investigated the structure of the footpoints of reconnected field lines. 
By comparing the distribution of reconstructed field lines and the observed flare ribbons, the trigger structure of the flare can be inferred. 
Our simulation suggests that the data-constrained simulation taking into account both the large-scale magnetic structure and the small-scale magnetic disturbance such as emerging fluxes is a good way to find out a flare productive active region for space weather prediction.
\end{abstract}


\section{INTRODUCTION}

Solar flare has been generally thought since a long time to be a result of the release of free magnetic energy contained in the active region \citep{gol60, par63, aly85}. The free energy can be stored as a result of the shear or twist of magnetic field near the polarity inversion lines (PIL) \citep{mor12, fal08}. It is also observed in many eruptive active regions that some sigmoidal structures formed across the active region before a flare or CME  occurred \citep{can99,gib06}. The sigmoidal structure basically shows that strong shear and twist exist in the active region. When an active region with high shear or twist occurs in the Sun, a small perturbation is likely to trigger the eruption of the sheared or twisted magnetic structures in this active region.

Several theories have been proposed to explain the triggering of solar flares. It is possible that the trigger process is related with converging flows \citep{inh92}, emerging flux \citep{hey77, chou98, lou15}, or reverse-shear magnetic field \citep{kus04}, which can affect the stability of the coronal magnetic field. Flare is also related to the formation and eruption of a large scale flux rope that can be caused by converging and shearing motion \citep{vanb89}, flux cancellation \citep{wang93}, or current carrying emerging flux \citep{wang94}. 

In the tether cutting scenario, reconnection of strongly sheared field below the magnetic arcades can trigger the eruption \citep{moor01}. \citet{kus12} proposed that two particular types of emerging fluxes can initiate the reconnection in the tether cutting scenario. On the other hand, \citet{ant99} proposed the magnetic breakout model where reconnection occurs due to the interaction of the  magnetic field with the overlying arcades at the null points \citep{aul00,sun13} or at bald patches \citep{wang02,jia12}. By using data-driven simulation, \citet{jia16} suggests that some jet-like reconnection can trigger the eruption, which corresponds to the breakout model. Moreover, according to magnetohydrodynamics (MHD) theory, solar flare can be thought to be triggered by the MHD instabilities, e.g., by torus instability \citep{kli06,dem10} or kink instability \citep{hood79,tor06} when the critical condition for the instability is obtained. These theories provide the mechanism for the free magnetic energy to be released as the kinetic energy and heat energy. 

The understanding of the flare trigger mechanism is crucially important to realize better prediction of when, where, and how flares will occur. However, in order to do that, one needs to be able to measure how stable the active region is, to determine whether it has enough free energy to be released, and to define the probability of flare. Therefore, it is necessary to study the flare trigger mechanism based on the observation and simulation.

Systematic studies of the flare trigger mechanisms have been performed by \citet{kus12}. They carried out ensemble MHD simulations with different Linear Force-Free Fields (LFFFs) and small bipole structures with different orientations imposed onto the LFFF. From their study, magnetic structures as well as the orientation of the bipoles which are effective in triggering a flare can be identified. They found that solar eruptions can occur as a result of both strong shear of the large-scale magnetic field near the PIL and the proper disturbance of the magnetic fields. The proper disturbances proposed in their study are the opposite polarity (OP) and the reversed shear (RS) structures. Opposite polarity refers to the small bipole structure whose polarity is opposite with respect to the polarity of the large-scale field structure at the polarity inversion line. Reversed shear polarity refers to the small bipole structure which is directed nearly opposite to the shear component of the field. In order to examine the model of the solar flare trigger mechanism proposed by \citet{kus12}, several observational analyses of flare events have been conducted by using Hinode \citep{kus12, bam13,tor13}, SDO \citep{bam14}, SOHO/MDI \citep{par13}, and New Solar Telescope (NST) data \citep{wang17}. From their results, several flare events can be explained to occur as a result of the flare trigger mechanism proposed in \citet{kus12}. 

However, the configurations of magnetic fields in the Sun are much more complex than the LFFF structures used in the study by \citet{kus12}. Due to the complexity of their structure, actual solar magnetic fields are very difficult to be reconstructed by this approach. For enabling this idea of solar flare trigger in the practical use of space weather forecasting, we need to use the concept of the solar flare trigger in more realistic coronal structure. For this purpose, here we study the flare trigger mechanism by \citet{kus12} with more realistic NLFFF magnetic field structures based on the observational data. Moreover, the goal of this study is to reveal which magnetic field configurations are effective for triggering a flare. Through this study, we aim at contributing to the improvement of flare prediction for space weather forecast.

Here we show and discuss the results of the MHD simulations for different configuration of small magnetic structures imposing in the NLFFF model of active region (AR) NOAA 10930 prior to the eruption of X3.4 flare in 2006 December 13. The NLFFF extrapolation method and MHD simulation scheme are described in Section 2. The results are presented in Section 3. In Section 4, we discuss and conclude how the reconstructed flare ribbon can be used to determine the flare trigger structure by comparing it with the observations. 

\section{NLFFF EXTRAPOLATION AND MHD SIMULATION SCHEME}

\subsection{Observations and Numerical Settings} \label{bozomath}
In this work, we used AR NOAA 10930 as a case study for our simulation. This active region was bipolar with the negative polarity spot larger than the positive polarity spot. It was very active since it produced at least 113 X-ray flares of different energy from 2006 December 4 to  2006 December 18 \citep{gop15}. Here, we focus on the X3.4 class solar flare which occurred at 02:14 UT on 2006 December 13. Many studies of this active region have been extensively conducted on various aspects, i.e. sheared field \citep{kub07, su07}, helicity and twist \citep{mag08, ino11, su09}, rotating sunspot \citep{min09, gop15}, NLFFF extrapolation \citep{sch08, ino12}, and MHD simulation \citep{fan11, amr14}. Sigmoidal structure has been reported to appear both from the observation as well as NLFFF extrapolation \citep{min09, ino12, amr14}.

We used vector magnetic field of AR 10930 derived from the Spectro Polarimeter (SP) data of the Solar Optical Telescope (SOT) instrument \citep{tsu08} on-board the Hinode satellite \citep{kos07} for the NLFFF extrapolation. We used Ca II H line (3968.5 $\text{\AA}$) data from the Broadband Filter Instrument (BFI) in the SOT for the purpose of examining how well the NLFFF and MHD simulation results agree with the structure of the flare ribbons. The X-ray image of the AR 10930 was obtained from the X-ray Telescope (XRT) on-board Hinode \citep{gol07}. This image is important to compare the NLFFF results with the coronal magnetic field configuration inferred from the X-ray image.

We inserted vector magnetogram data obtained from the Hinode/SP as a bottom boundary condition from the original $1000 \times 512$ pixels in order to fit to the $240 \times 128 \times 128$ uniform grid used in the simulation box. The magnetogram's field-of-view is $297 \times 163$ arcsec, corresponding to $214 \times 118$ Mm on the Sun. The simulation box represents the rectangular domain of $(-0.5L,-0,25L,0) \leq (x,y,z) \leq (0.5L,0.25L,0.5L)$, where \textit{L} is the normalization of the spatial length, which has the actual value of about 214 Mm. 

\subsection{NLFFF Method} \label{bozomath}
We follow the MHD relaxation method of \citet{ino14a} to reconstruct the coronal field of the active region we are interested in. We use vector magnetic field data obtained from the Hinode/SP magnetogram at 20:30 UT on 2006 December 12, which was about six hours before the X3.4 flare onset. The potential field of the active region is calculated as an initial condition from the normal component $B_{z}$ of the vector magnetic field on the photosphere by using the Fourier method \citep{ali81}. The initial density is chosen to be uniform. After inserting the observed tangential components ($B_{x}$ and $B_{y}$) into the bottom boundary, the magnetic field in the whole domain is then evolved towards the force-free state. This evolution process is governed by the set of equations for zero plasma beta,

\begin{equation}
\rho = \rho_0 \frac{|{\bf B}|}{B_0}
\end{equation}

\begin{equation}
\frac{\partial \textbf{\textit{v}}}{\partial t}=-(\textbf{\textit{v}}\cdot\bm{\nabla})\textbf{\textit{v}}+\frac{1}{\rho}\textbf{\textit{J}}\times\textbf{\textit{B}}+\nu\bm{\nabla}^{2}\textbf{\textit{v}} , 
\end{equation}

\begin{equation}
\frac{\partial \textbf{\textit{B}}}{\partial t}=-\bm{\nabla}\times(-\textbf{\textit{v}}\times\textbf{\textit{B}})+\eta_{_{NLFF}}\bm{\nabla}^{2}\textbf{\textit{B}}-\nabla\phi ,
\end{equation}

\begin{equation}
\frac{\partial \phi}{\partial t}+c_{h}^{2}\bm\nabla\cdot\textbf{\textit{B}}=-\frac{c_{h}^{2}}{c_{p}^{2}}\phi ,
\end{equation}

\begin{equation}
\textbf{\textit{J}}=\bm\nabla\times\textbf{\textit{B}} ,
\end{equation}
where $\rho$ is the plasma density, \textbf{\textit{v}} is the plasma velocity, \textbf{\textit{J}} is the current density, and \textbf{\textit{B}} is the magnetic flux density. In this method, Equation (1) defines a pseudo-density ($\rho$), which is proportional to $\mid\textbf{\textit{B}}\mid$, in order to ease the relaxation by maintaining the Alfv$\acute{e}$n speed in space \citep{ino13}. Equation (2) is the equation of motion for the zero plasma beta condition neglecting gravity. The last term in the induction equation (3) includes the $\bm\nabla\cdot \bm{B}$ cleaning potential ($\phi$). The cleaning potential equation (4) was introduced by \citet{ded02} to reduce deviation from the solenoidal condition $\bm\nabla\cdot\textbf{\textit{B}}=0$, where $c_{h}$ and $c_{p}$ are the coefficients related to advection and diffusion of $\bm\nabla\cdot \bm{B}$, respectively. 

The magnetic field (\textbf{\textit{B}}) in the calculation is normalized by $B_{0}$, which equals to 4000 G. Velocity, time, and electric current density are normalized by $V_{A}  \equiv B_{0}/(\mu_{0} \rho_{0})^{(1/2)}$, $\tau_{A}\equiv L/V_{A}$, and $J_{0}=B_{0}/\mu_{0} L$, respectively. In the typical AR, $\rho_{0}=1.67 \times 10$\textsuperscript{-12} kg/m\textsuperscript{3}, so that $V_{A} \approx 275$ Mm/s, $\tau_{A} \approx 0.8$ s, and $J_{0}\approx 1.5$ mA/m\textsuperscript{2}. We set the coefficients following \citet{ino14a}, where $c_{p}^{2}$ and $c_{h}^{2}$ are 0.1 and 0.04, respectively. The non-dimensional viscosity ($\nu$) in equation (2) is set as a constant ($1.0 \times 10^{-3}$ ). Magnetic diffusivity ($\eta$) in equation (3) is defined as
\begin{equation}
\eta_{_{NLFF}}=\eta_{_{0}}+\eta_{_{1}}\frac{\mid\textbf{\textit{J}}\times\textbf{\textit{B}}\mid\mid\textbf{\textit{v}}\mid^{2}}{\mid\textbf{\textit{B}}\mid^{2}} ,
\end{equation}
where $\eta_{_{0}}=5.0\times10^{-4}$  and $\eta_{_{1}}=1.0\times10^{-3}$ are non-dimensional parameters in the units of $\mu_0 V_A L$ and $(\mu_0 L)^2 / V_A$, 
respectively.

At the bottom boundary, once we run the program, the tangential components from the potential field are incrementally changed into the observed tangential components while all physical values in the other boundaries are fixed. After the bottom boundary values of magnetic vector field are completely changed into the observed values, we set all the physical values for all boundaries, including the bottom boundary to be fixed during the calculation. The method for the NLFFF extrapolation and parameter setting in this work are almost identical to the NLFFF method by \citet{ino14b}.

\subsection{Numerical Scheme for the MHD Simulation} \label{bozomath}
The MHD simulation is performed in the same grid as the NLFFF extrapolation. It uses the NLFFF model and the corresponding density as the initial conditions. The non-ideal zero-beta MHD equations are solved in the MHD simulation. Hence, the induction equation now takes the form
\begin{equation}
\frac{\partial \textbf{\textit{B}}}{\partial t}=-\bm{\nabla}\times(-\textbf{\textit{v}}\times\textbf{\textit{B}}+\eta_{_{MHD}}\textbf{\textit{J}}) ,
\end{equation}
and the continuity equation, 
\begin{equation}
\frac{\partial \rho}{\partial t}=-\bm\nabla\cdot(\rho\textbf{\textit{v})},
\end{equation}
replaces equation (1).
The magnetic diffusion ($\eta$) in equation (7) is defined as an anomalous resistivity following \citet{ino14b},
\begin{equation}
    \eta_{_{MHD}}(t)= 
\begin{cases}
    \eta_{_{2}},        & J\leq j_{c}\\
    \eta_{_{2}}+\eta_{_{3}}\left(\frac{J-j_{c}}{j_{c}}\right),              & J > j_{c},\\
\end{cases}
\end{equation}
where $\eta_{_{2}}=1.0\times 10^{-5}$, $\eta_{_{3}}=5.0\times 10^{-3}$, and the threshold current density, $j_{c}=300$. This anomalous resistivity can be expected to enhance the reconnection of the field lines in the regions of strong current \citep{ino14b}.

\citet{kus12} suggested that the trigger structure is located near the photospheric polarity inversion line (PIL). Accordingly, we expect the area near the PIL of the core field to be particularly effective for triggering a flare. According to \citet{bam13}, the trigger structure of the X3.4 flare studied here was situated in the area marked by the yellow circle in Fig.\ 1(a). They showed that a highly sheared structure existed along the PIL and that a small positive polarity magnetic island grew near the PIL as is shown in Fig.\ 1(b). This location was obtained from their study of the topological features of the flare ribbons and their associated highly sheared structure. They found that the emerging flux of the magnetic island, which was located between the flare ribbons, triggered the X3.4 flare six hours after this magnetogram was taken. The orientation of the bipole flux was opposite to the orientation of the large-scale magnetic field of the active region, and thus it could lead to the eruption of the sheared or twisted magnetic field lines by introducing the reconnection which formed and destabilized the flux rope. Therefore, we chose this location as the place where the small bipole structure is injected as flux that emerges into the initial field in our simulation to trigger the eruption.

The emerging flux model follows the method of \citet{kus12}, where the small bipole is made from a magnetic torus that ascends from below the simulation box. The bipole structure is a sphere with radius $r_{e}$ filled with a purely toroidal field of uniform strength, $B_{e}$. An electric field $\bm B_{e}\times\bm v_{e}$ is imposed in the cross-section of the bottom plane to let the torus ascend with velocity $v_{e}$, chosen to be constant during the period $0\leq t\leq \tau_{e}(=r_{e}/v_{e})$. The injected bipole structure has the azimuthal orientation angle, $\phi_{e}$, defined as shown in Fig.\ 1(c). The bipole is injected at the coordinate $P(x=294,y=-98)$ arcsec and has a magnetic intensity, $B_{e}=15$ and a radius, $r_{e}=0.01$. It starts to ascend with the constant velocity $v_{e}=0.02$ at $t=0$, and is stopped at $t=\tau_{e}=0.5$ when the center of the sphere reaches the bottom plane. This velocity is higher than the typically observed photospheric velocities, but still slower than the coronal Alfv$\acute{e}$n velocity and, therefore, appropriate for the problem studied here. We perform simulations with various angles $\phi_{e}$. Eight cases are run as summarized in Table 1. Case C ($\phi_{e}=110^\circ$) is displayed in Fig.\ 1(d). 

Based on its orientation with respect to the pre-existing field, the bipole configurations imposed in our simulations can be classified as right polarity ($\phi_{e}\approx 0^\circ$), reversed shear ($\phi_{e}\approx 90^\circ$), opposite polarity ($\phi_{e}\approx 180^\circ$), and normal shear ($\phi_{e}\approx 270^\circ$) type, using the terms introduced by \citet{kus12}. RS type configuration is defined as the bipole flux whose orientation is almost oppositely directed to the shear (non-potential) field component. Here we define the RS-type to be the bipole with $\phi_{e}\approx 90^\circ$ because the shear field in the area around the PIL has left-handed twist so that the magnetic helicity is negative. The left-handed shear and twist can be seen from the reverse S shape of the sigmoid and from the angle between the threads of the sigmoid and the PIL. This was confirmed by a computation of twist map by \citet{ino12}. OP type, on the other hand, is defined as the bipole structure with the orientation almost opposite to the averaged potential field. 

The constraint for the tangential components of magnetic field on the top and bottom boundaries is set to be released during the simulations, whereas the normal components are fixed except for the area where the bipole flux is emerging. At the side boundaries, all physical values are fixed during the simulations. Due to the relatively small size of the numerical box and the fixed side boundary conditions, we cannot expect that the simulation will produce a large expansion of the field such as coronal mass ejection. We only focus on the dynamics of the beginning phase of the flare process.

\section{RESULTS}
\subsection{NLFFF Extrapolation} \label{bozomath}
The top view of field lines in the NLFFF model, plotted over normal component of the magnetogram data, is shown in Fig.\ 2(a). It shows that open magnetic field dominates the active region in the area within and surrounding the negative polarity. This is due to the imbalance of the flux between the negative and positive polarities in the active region. The coronal magnetic field is closed in the area surrounding the polarity inversion line. We call this the core field of the active region. As shown in Fig.\ 2(a), the core field shows a strong shear, which can also be seen in the photosphere (Fig.\ 1(b) and (d)).  

The extrapolated NLFFF is strongly sheared particularly on the lower part of the corona. This core field may contain a large amount of free energy, since it differs strongly from the potential field. The comparison with X-ray image taken by the XRT instrument onboard Hinode (Fig.\ 2(b)) shows that the NLFFF model agrees well with the observation in terms of the presence of high shear at the PIL. Moreover, the NLFFF model infers that the sigmoidal structure at the PIL consists of short arcade-type field lines (Fig.\ 2(a)). This sigmoidal structure is important because it shows that the core field of the AR is highly sheared (Su et al. 2007; Min \& Chae 2009). The reverse-S shape is well reproduced in the NLFFF extrapolation used here as well as in several previous works by \citet{ino12} and \citet{amr14}. 

\subsection{MHD Simulation} \label{bozomath}

As a reference case, we first carry out a simulation without imposing any external perturbation. This simulation is performed to show the nature of the system if there is no emerging flux imposed. It is verified that the residual Lorentz force in the NLFFF model is too weak for triggering an eruption (Fig.\ 3(a) and (b)). The simulation shows that the shear of the magnetic field slightly weakens under the condition of the released tangential field components in the bottom boundary, as shown in Fig.\ 3(b). It is easy to understand that the magnetic field will naturally relax to a lower-energy state which is toward the potential field configuration. This verifies that any eruption of the NLFFF must be driven by an external disturbance. The relaxation to a stable equilibrium state can also be seen from the evolution of the kinetic energy in the box in Fig.\ 3(c), where a brief initial rise (due to the residual Lorentz force in the initial condition) is followed by a monotonic decrease (similar to Run B in \citet{ino14b}). 

We find that several configurations of the trigger structure can lead to an eruption; these are the structures in cases C, D, E, and F. However, each type of triggering structure creates a different dynamics and topology of the erupting flux rope. Here we carefully analyze the eruption of the flux ropes in our simulations to clarify the typical dynamics involved in the erupting process. Based on the relation between flare reconnection and flux rope formation, all eruptive cases in our simulation can be categorized into two distinct groups, which are ``eruption-induced reconnections" and ``reconnection-induced eruptions". The former is the case when the flux rope is formed before the flare reconnection occurs. In this case, the emerging bipole flux triggers the creation of an unstable flux rope through pre-flare reconnection. Subsequently, the flare reconnection is generated below the flux rope during its eruption. As for the latter, the role of the emerging bipole is to trigger the reconnection between pre-existing magnetic field  by reducing the shear of the overlying field which then creates an unstable flux rope. These two types of dynamical process were also observed in the simulations conducted by \citet{kus12}.       

Eruption-induced reconnection features are clearly observed in the simulation results for Case E ($\phi_{e}=180^\circ$) and Case F ($\phi_e=225^\circ$). However, although they share some common features of eruption-induced reconnection, each case has its own characteristics of the topological structure, due to the difference in the azimuthal angle of the bipole. Figs.\ 4(a)-(d) show the evolution of the eruption-induced reconnection from the initial condition until the expansion of sigmoidal flux ropes. At the beginning (Fig.\ 4(a)), the magnetic field lines do not form any large twisted flux rope. After the small bipole structure emerges in the photosphere, it starts to reconnect with the pre-existing field and forms a flux rope with sigmoidal shape (Fig.\ 4(b)). The flux rope, then acquires a higher twist (Fig.\ 4(c)) and expands outward (Fig.\ 4(d)). Figs.\ 5(a)-(b) show the detailed dynamics of the eruption-induced reconnection seen from a different point of view. When the bipole flux emerges, a flux rope occurs immediately after the reconnection of the magnetic field near the PIL via the OP-type structure, and high electric current regions (shown by red shade) are formed (Fig.\ 5(b)). This reconnection tends to create a large flux rope with high twist (yellow lines in Fig.\ 5(c)) that quickly erupts, as can be seen in Fig.\ 5(d), where the flux rope lifts the overlying field and finally induces flare reconnection below the flux rope. 

In Cases C and D, the process of the eruption follows reconnection-induced eruption scenarios found in \citet{kus12}. Figs.\ 6(a)-(d) show the evolution of the reconnection-induced eruption for Case C. At the first step, the core magnetic field near the PIL reconnects with the RS structure bipole (Fig.\ 6(b)). Subsequently, the overlying field at the center of the RS structure collapses (Fig.\ 6(c)) and starts to create a large flux rope (Fig.\ 6(d)). Figs.\ 7(a)-(b) show the detailed dynamics of flux rope formation in the reconnection-induced eruption seen from a different point of view. Just after the bipole emerges, some of the pre-existing field lines near the PIL (blue lines in Fig.\ 7(b)) are in contact with the bipole field and create a current sheet. This reconnection reduces the sheared field (blue lines), which causes the overlying field to collapse toward the center. Part of the collapsed field finally reconnects with the bipole structure (green lines in Fig.\ 7(c)-(d)). However, the higher overlying field tends to form a flux rope, which then erupts upward, as shown by the yellow lines in Fig.\ 7(d). The topological structure of these steps in the present simulation is consistent with the previous simulation by \citet{kus12} although it is more difficult to observe compared to the previous simulation. This is due to the more complex configuration of the real coronal magnetic field rather than a symmetric structure by the initial boundary condition of LFFF in the \citet{kus12} simulation. All of these steps are also observed in Case D.

From the simulation results we find that the azimuthal angle of the bipole structure plays a very important role in determining the overall dynamics of the magnetic field. Some imposed bipole structures clearly are not effective in triggering a flare, while some others are very effective in triggering a flare, with some aspects of the evolution also depending on the azimuth angle. A summary of the simulation results is shown in Fig.\ 8. It is found that the events that do not show any eruptive characteristics are the cases where the imposed emerging flux is oriented relatively parallel to the potential or shear (non-potential) components of the pre-existing magnetic core field at PIL. In our simulations, these are the cases A, B, G, and H with $\phi_{e}=10^\circ$, $50^\circ$, $270^\circ$, and $315^\circ$, respectively. Otherwise, an eruption is triggered.

\subsection{Comparison With Observation} \label{bozomath}
Flare ribbons can well represent the topology of the reconnecting magnetic field, so that they can be used to check the results of the simulations. The ribbons mark the footpoints of magnetic field lines that reconnect during the flare. A proxy for flare ribbons in the simulations is made by following the method introduced by \citet{tor13} and applied also in \citet{ino14b}. We calculate the total displacement of the footpoint for each field line for a given time and consider field lines with a large footpoint displacement to be reconnected ones. We trace each magnetic field line from each point $({\bf{x_0}})$ in the bottom plane and identify the end point of the field line ${\bf{x_1}} ({\bf{x_0}},t_0)$. Here, the end point position ${\bf{x_1}}$  as a function of start point ${\bf{x_0}}$ and time $t_n$ is denoted as ${\bf{x_1}} ({\bf{x_0}},t_n$). After some time, we trace again the end point of the field line $({\bf{x_1}} ({\bf{x_0}},t_n))$ from the start point $({\bf{x_0}})$. Therefore, the displacement of the end point position for one start point ${\bf{x_0}}$ is given by 
\begin{equation}
\delta({\bf {x_0}},t_{n})={|{\bf {x_1}}({\bf {x_0}},{t_{n+1}})-{\bf {x_1}}({\bf {x_0}},{t_n})|}.	
\end{equation}
By integrating $\delta$ for a given time, $t=t_{N}$, we can obtain the total displacement of the end point from the initial state to an arbitrary time step,
\begin{equation}
\Delta x(x_{0},t)=\sum_{n=0}^{N} \delta(x_{0},t_{n}).
\end{equation}
We assume that the high value of total displacement ($\Delta x$) is due to reconnection. Synthetic flare ribbons are constructed by plotting the value of the total displacement of all field lines as a function of their footpoint position on the bottom plane. The results are shown in Fig.\ 9 for the eight different cases. In order to emphasize the distribution of the flare ribbons, we only plot footpoints with the total displacement exceeding 0.04. We also calculate the total reconnected flux of $B_z$ from the areas within the red square in Fig.\ 1(a) that show a large displacement of field lines and denoted this as $\Phi_{rec}$ in Table 1. These total fluxes show how much flux reconnects when the emerging flux meets the condition of a flare trigger.       

It is obvious from Fig.\ 9 that the flare ribbons constructed from the simulations are different in each case. This suggests that the topology of the reconnected magnetic field is unique for each case. Therefore, the dynamics of the magnetic field due to the interaction between the pre-existing magnetic field and the emerged bipole structure depends considerable on the orientation of the bipole. The results also show that the topology of the magnetic field involved in the eruption process triggered by the RS and OP type structures differs, even though both cases end up with eruption. The RS-type structure (Case C and Case D) tends to produce local flare ribbons in the area of the core-field of the active region. On the other hand, the OP-type structure (Case E) can generate more extended flare ribbons of more complex structure.  

\section{DISCUSSION AND CONCLUSION}

From the results of all cases in our simulation, we can classify each case of the simulation in three different categories: non-eruptive events, reconnection-induced eruption events, and eruption-induced reconnection events. In the non-eruptive cases, the small imposed bipole structure is relatively parallel to the potential or shear components of the average vector magnetic field of the large bipolar spots of AR 10930. It can be easily understood that in such kind of configurations, magnetic reconnection is very difficult to occur. Therefore, the small amount of magnetic flux involved in the reconnection is clearly responsible for the non-eruptive behavior. 

Our results suggest that the effective structures for triggering solar flares are the opposite polarity (OP) and the reversed shear (RS) configurations imposed on a magnetic field with strong shear. Our simulations indicate that the both types of configuration can trigger the eruption of the sheared field, although they show different erupting mechanisms (see Figs.\ 5(a-d) and Figs.\ 7(a-d)). The difference between them is in the causality of onset process of solar flare and solar eruption. The synergetic interaction between the reconnection and the eruption must be the main driver of large flares. However, just in the beginning, one process has to cause another process in order to initiate the mutual interaction. In the eruption-induced reconnection, flux rope becomes unstable and erupts before the flare reconnection starts. On the other hand, in the reconnection-induced eruption, the flare reconnection starts first, and reconnection generates flux rope, which becomes unstable and erupts.

In the eruption-induced reconnection, the pre-flare reconnection starts just above the photosphere before the flux rope is launched from the chromosphere. It results in the propagation of ribbons from the center of flare to the location where two-ribbon appears in the main phase of flare. This type of propagation of ribbons in pre-flare brightening was found by \citet{kus12}, and recently \citet{wang17} successfully observed the detail structure and dynamics of the pre-flare using New Solar Telescope. The results are well consistent with the model of the eruption-induced reconnection. On the other hand, the reconnection in the reconnection-induced eruption process starts on some portion in the corona. The first ribbons should appear as the separated two-ribbons and the propagation from the center cannot appear. This type of flare was also found by \citet{kus12} and \citet{bam17}. Moreover, the comparison between the total magnetic fluxes involved in the reconnection (Table 1) shows that the OP-type and RS-type structures tend to involve larger fluxes in the reconnection compared to the non-eruptive cases.    

The comparison between synthetic and observed flare ribbons suggests that not all the cases of simulations in our simulation agree with the observation. It can be qualitatively seen that only Case E can closely reproduce the shape and location of the flare ribbons. Fig.\ 10(a) shows the flare ribbons reconstructed in Case E plotted over a Ca II H image taken by the Hinode/SOT instrument. From this image, although the reconstructed flare ribbons cannot perfectly agree with the observation, the main features of the ribbons are reproduced. The simulation of Case E does find ribbons that extend far out of the core region, but their details differ from the observation. This is due to the limited size of the computational box which cannot include the magnetic field far from the core active region and its connectivity with distant structures. 

In Fig.\ 10(b), we additionally plot the field lines associated with the synthetic flare ribbons in Case E. It is found that most of the field lines traced from the reconstructed flare ribbon belong to the flux rope formed by the reconnection between the OP-type emerging flux and the pre-existing sheared field. This result is consistent with the result of \citet{kus12} and \citet{bam13}, who studied the trigger structure through observational analysis of the magnetic field from magnetogram data and Ca II H flare ribbons. In the latter study, it was pointed out that the OP-type structure might be associated with emerging flux and that the reconnection of the OP-type emerging flux with the pre-existing sheared field may create a flux rope. Based on this assumption, it was concluded that the main flare ribbon structures both in the east and west side of the PIL were located at the feet of a twisted flux rope formed by reconnection. This main flare ribbon structures can be explained by the simulation by \citet{kus12}. The flare ribbons in our simulation also correspond to the footpoints of the flux rope denoted as $F-F^\prime$ in their simulation (see Fig.\ 3 in \citet{kus12}). 

Our simulations show that the trigger mechanism proposed by \citet{kus12} can be applied to the real coronal-like magnetic field environment. Although the trigger structures appear as emerging flux in our simulations, it is possible that the trigger may come from other processes as long as the configuration of the trigger structure exists in the proper way. Several possible ways are a splitting of the sunspot that may lead to a flow towards the PIL \citep{lou14} and a series of bipolar emergence \citep{tor13}. \citet{kur02} showed that such a configuration indeed occurred in a flare-productive active region and could be explained by an emerging twisted flux rope. This emerging twisted flux rope evolved and appeared as a sunspot motion or rotation by means of the kink instability.   

Finally, in this study, we succeed to confirm the OP-type trigger structure and mechanism responsible for X3.4 flare in AR 10930, which were proposed by \citet{kus12} based on observation and simulations using an idealized AR model. Although we run the simulations under several constraints of the limited size of the simulation box and time scale, we demonstrate that MHD simulations can be a powerful tool to examine the trigger process of flares. This study can be important for space weather prediction, especially for the method that rely more on the physics-based approach rather than the statistical approach. Some future work to examine the critical size and location of the emerging flux that can trigger an eruption will be conducted elsewhere in the near future.

\acknowledgments
This work was supported by JSPS/MEXT KAKENHI Grant Nos. 23340045 and 15H05- 814 and by MEXT as ”Exploratory Challenge on Post-K computer” (Elucidation of the Birth of Exoplanets [Second Earth] and the Environmental Variations of Planets in the Solar System). JM is grateful to Indonesia Endowment Fund for Education (LPDP) for supporting his stay and study at Nagoya University. The authors would like to thank Dr. Bernhard Kliem for his valuable comments that improved the manuscript. The authors are grateful to the anonymous referee for the constructive comments which lead to the improvement of the manuscript. Hinode is a Japanese mission developed and launched by ISAS/JAXA, with NAOJ as domestic partner and NASA and STFC (UK) as international partners. It is operated by these agencies in co-operation with ESA and NSC (Norway). The ambiguity resolution code used herein was developed by K.\ D.\ Leka, G.\ Barnes, and A.\ Crouch with NWRA support from SAO under NASA NNM07AB07C. Visualizations of NLFFF and simulation fields were produced by VAPOR \citep{cly07,cly05}.  

\clearpage

\appendix

\clearpage

\begin{figure}
\epsscale{1.0}
\plotone{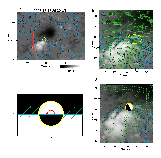}
\caption{(a)Distribution of the normal component of AR 10930
vector magnetic field, $B_{z}$, on the bottom of the simulation box. White and black is positive and negative polarity, respectively. Blue contour is the polarity
inversion line and yellow circle marks the area where the bipole field is injected. 
(b)Vector magnetic field map obtained from Hinode/SP magnetogram data on 2006-12-12 20:30 UT overplotted on the normal component of vector magnetic field for the area within the red box in (a). The dashed yellow ellipse shows the presence of OP-type magnetic island. (c)Orientation of the azimuthal angle $\phi_{e}$ of the emerging flux (bipole field) on the X-Y plane as seen 
in the top view of the simulation box. Green arrows represent the background transverse magnetic field. (d)Enlarged view of the red box in (a) with the vector magnetic field map for the binned data used in the bottom boundary of the simulation box. The yellow circle shows the orientation of the imposed bipole flux when it stops to ascend in the simulation run with $\phi_{e}=110^\circ$.       
\label{fig1}}
\end{figure}

\clearpage

\begin{figure}
\epsscale{1.0}
\plotone{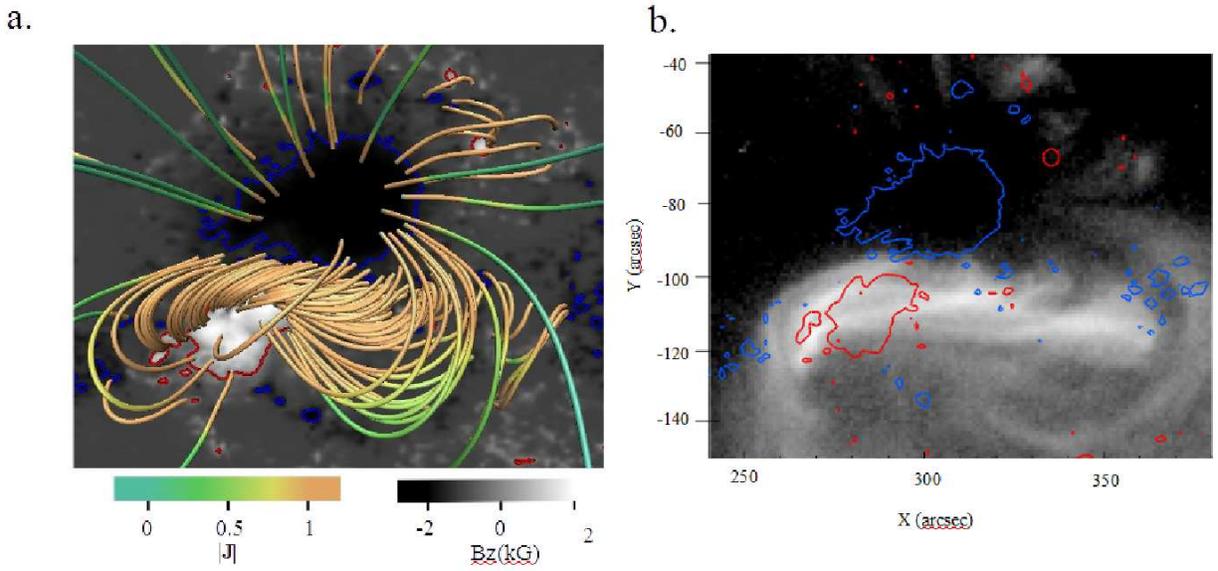}
\caption{(a) NLFFF model of AR 10930 overplotted on the background image of 
$B_{z}$, with the blue and red contours showing the -800 G and 800 G levels of 
the $B_{z}$, respectively. Field lines are plotted with a color representing current density. The field lines with strong current density form a
sigmoidal pattern. (b) The same contours of $B_{z}$ plotted on the X-ray image
observed by Hinode/XRT on 21.00 UT, which shows a sigmoidal structure that corresponds
to the core-field in the NLFFF model.
\label{fig2}}
\end{figure}

\clearpage

\begin{figure}
\plotone{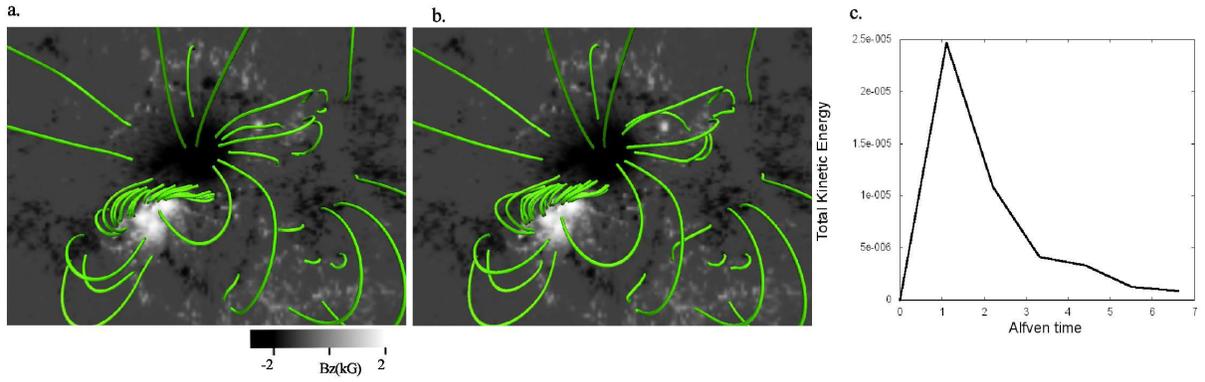}
\caption{(a)Magnetic field of AR 10930 in the same area as Fig.\ 2(a) for the simulation without emerging flux at (a)t=0 and (b)t=3.3. The kinetic energy in the box is plotted in (c).}
\end{figure}

\clearpage

\begin{figure}
\epsscale{0.7}
\plotone{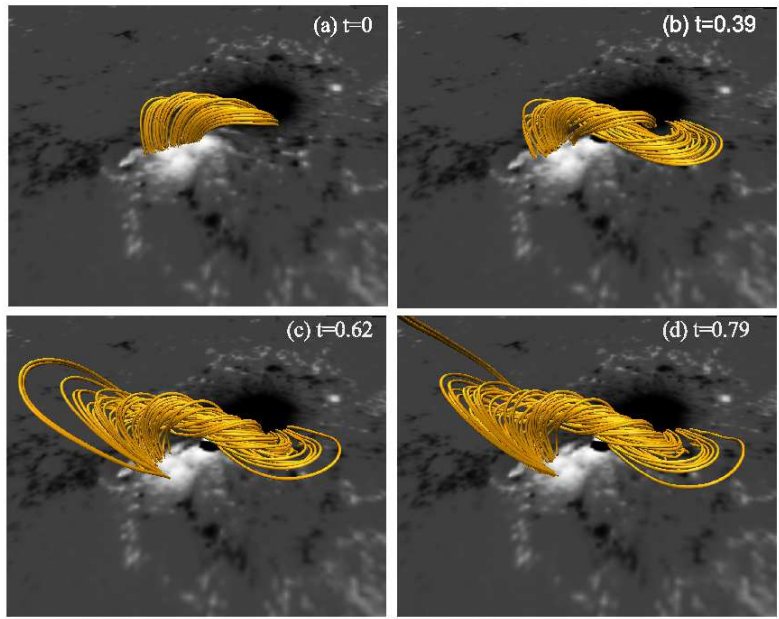}
\caption{(a)-(d) Bird eye view of the time evolution of the core magnetic field (gold lines) for Case E with imposed bipole azimuth angle $\phi_{e}$=$180^{0}$.} 
\end{figure}

\clearpage

\begin{figure}
\epsscale{0.7}
\plotone{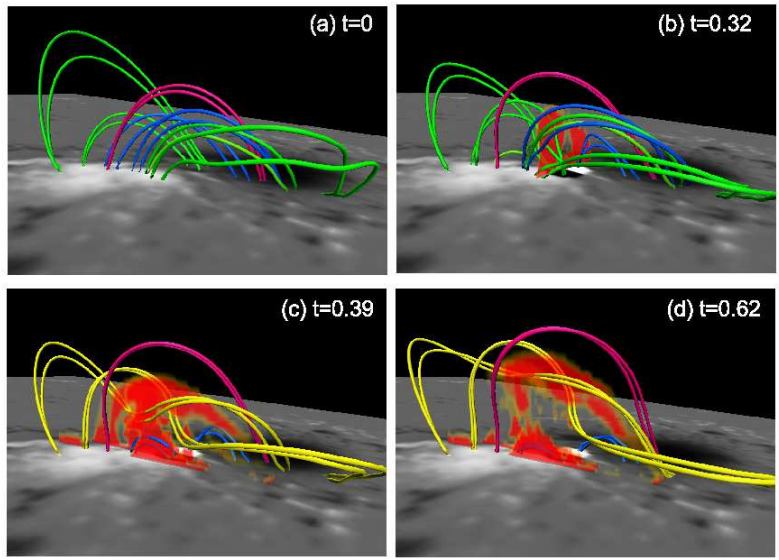}
\caption{Dynamics of eruption-induced reconnection caused by the OP-type structure of emerging flux for Case E: (a) the initial state, (b) after the bipole flux is injected, (c) after the flux rope starts to appear, and (d) when the flux rope erupts. Green lines show the field lines before the flux rope is formed by reconnection with the OP-type bipole field. Blue lines show the magnetic field lines which changed their connectivity due to the reconnection with the imposed OP-type structure. Purple lines show the magnetic field lines which retain the same connectivity. Yellow lines show the created flux-rope due to the reconnection between green lines in (a) and (b). The red areas show enhanced current density with ${|{\bf J}|}> 30$. (An animation of this figure is available.)}
\end{figure}

\clearpage

\begin{figure}
\epsscale{0.7}
\plotone{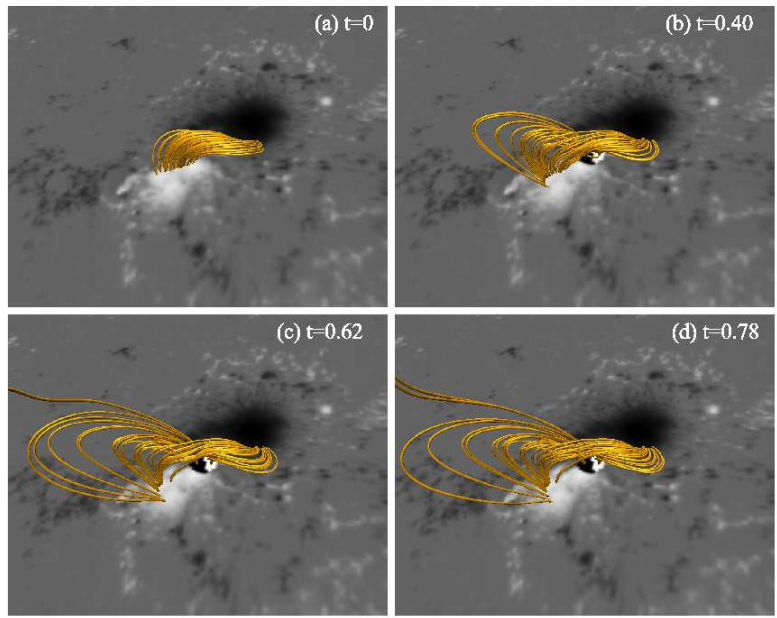}
\caption{(a)-(d) Bird eye view of the time evolution of the core magnetic field for Case C with imposed bipole azimuth angle $\phi_{e}=110^{0}$.} 
\end{figure}

\clearpage

\begin{figure}
\epsscale{0.7}
\plotone{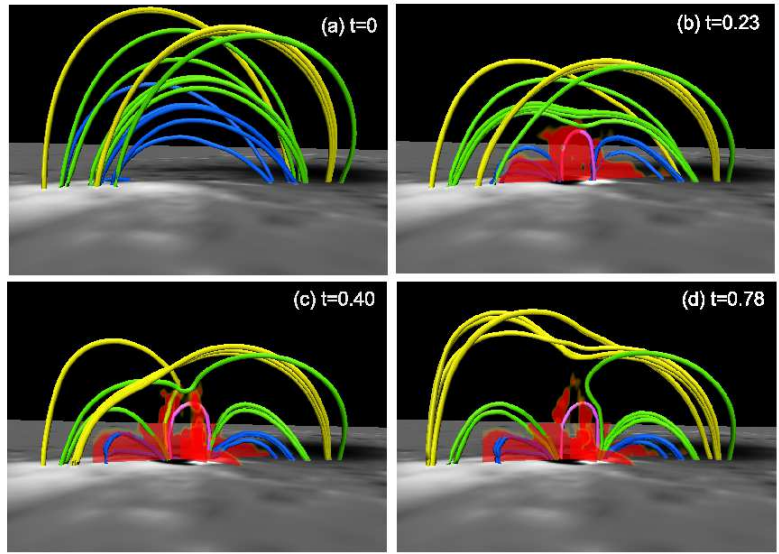}
\caption{Dynamics of reconnection-induced eruption caused by RS-type structure for Case C: (a) the initial state, (b) after bipole flux is injected, (c) during the formation of the flux rope, and (d) when the flux rope erupts. Blue lines show the magnetic field lines which changed the connectivity due to the reconnection with the imposed RS-type structure. Green lines show the magnetic field lines which collapsed to the center of the RS-type structure and then finally reconnected with the imposed RS-type structure. Yellow lines show the magnetic field lines that created the flux rope. Red areas correspond to intense current density layers with ${|{\bf J}|}> 40$. (An animation of this figure is available.)}
\end{figure}

\clearpage

\begin{figure}
\epsscale{0.9}
\plotone{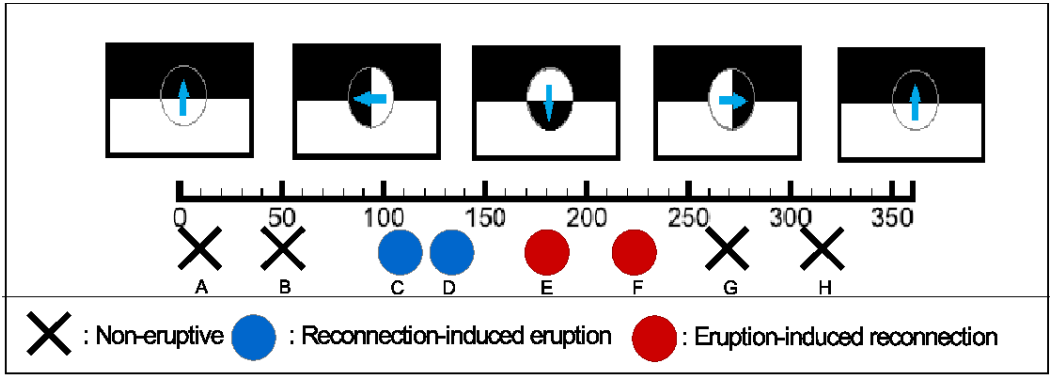}
\caption{Summary of the simulations for Cases A-H and the classification of the events based on the eruptive behavior and trigger structure. The upper panels show the orientation of the emerging bipole structure and the corresponding azimuth angle.}
\end{figure}

\clearpage

\begin{figure}
\epsscale{0.8}
\plotone{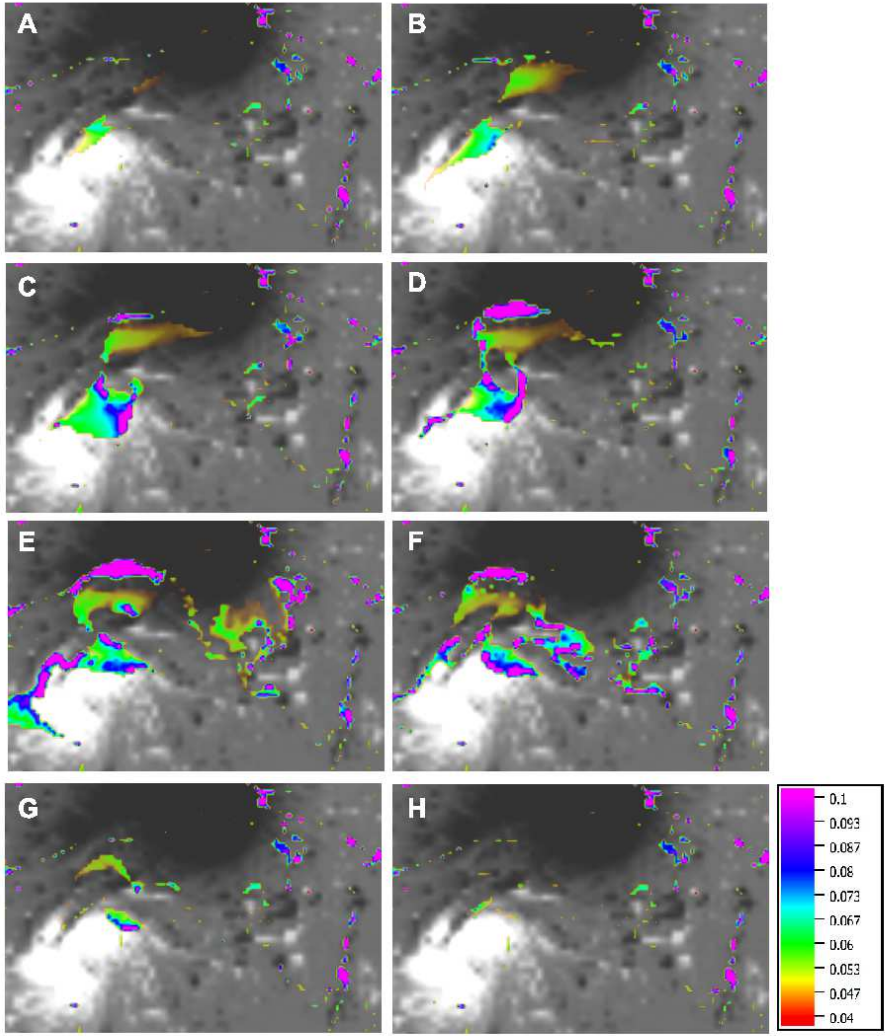}
\caption{Synthetic flare ribbons constructed from the simulations at t=0.5, plotted over the normal component of the magnetic field. The color bar shows the total displacement of the field line footpoints derived from equation 11.}
\end{figure}

\clearpage

\begin{figure}
\epsscale{1.0}
\plotone{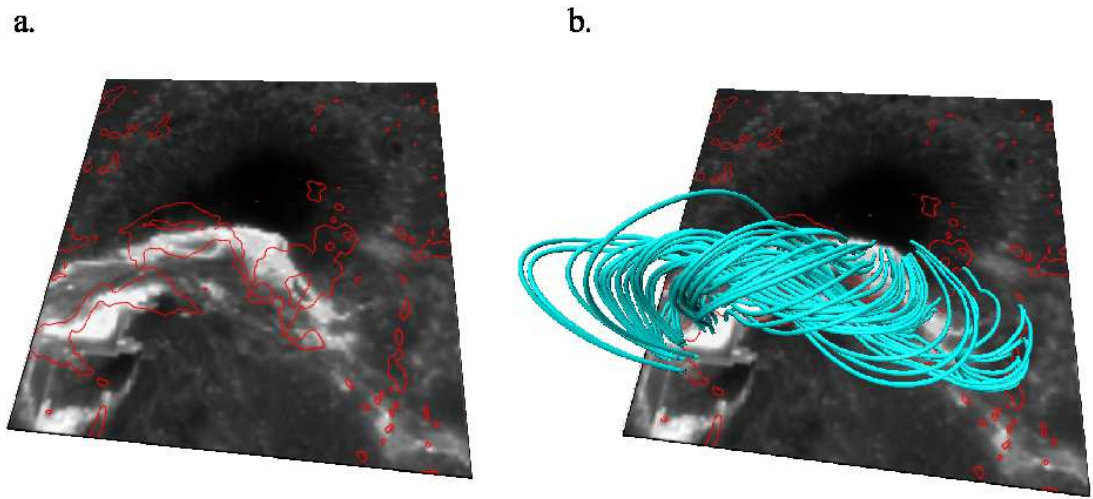}
\caption{(a) Bird eye view of synthetic flare ribbons from Case E (red contour) at t=0.5, plotted over the flare ribbons of the Ca II H line from the Hinode/SOT observation. Red contours mark the total footpoint displacement of 0.04 in the simulation. (b)Field lines of the erupted field plotted over the panel (a).}
\end{figure}

\clearpage

\clearpage

\begin{table}
\begin{center}
\caption{Azimuth angle ($\phi_{e}$) and total reconnected flux ($\Phi_{rec}$) in the area of the red square in Fig.\ 1 for the different cases performed in the simulations, estimated from the flux covered by field lines with a large displacement $\Delta x(x_0)$ .\label{tbl-1}}

\begin{tabular}{cccc}
\tableline\tableline
Run Case & Orientation ($\phi_{e}$) & $\Phi_{rec}$ \tablenotemark{a}\\
\tableline
$A$ &$10^{0}$ &$2.44 \times 10^{-4}$\\
$B$ &$50^{0}$  &$4.68 \times 10^{-4}$\\
$C$ &$110^{0}$ &$6.2  \times 10^{-4}$\\
$D$ &$135^{0}$ &$6.27 \times 10^{-4}$\\
$E$ &$180^{0}$ &$8.12 \times 10^{-4}$\\
$F$ &$225^{0}$ &$4.81 \times 10^{-4}$\\
$G$ &$270^{0}$ &$2.14 \times 10^{-4}$\\
$H$ &$315^{0}$ &$1.29 \times 10^{-4}$\\
\tableline
\end{tabular}
\tablenotetext{a}{The values are normalized by $\Phi_{0} \equiv B_0 L^{2} = 1.83 \times 10^{24}$ Mx}

\end{center}
\end{table}

\clearpage

%% The following command ends the manuscript. 

\end{document}